\documentclass{appolb}

\usepackage{graphicx}
\usepackage{rotating}
\usepackage{amsmath}
\usepackage{bm,amsfonts,amssymb}

\def\be{\begin{equation}}
\def\ee{\end{equation}}
\def\bea{\begin{eqnarray}}
\def\eea{\end{eqnarray}}
\def\bdm{\begin{displaymath}}
\def\edm{\end{displaymath}}
\def\upr{\uparrow}
\def\dar{\downarrow}
\newcommand{\barr}{\begin{array}}
\newcommand{\earr}{\end{array}}
\newcommand{\ttpo}{$2\,{}^{3\!}P_1$}
\newcommand{\tpo}{${}^{3\!}P_1$}
\newcommand{\dsp}{\displaystyle}
\begin{document}
\title{Is the $X(3872)$ a molecule?
\thanks{Presented by S. Coito at the Workshop ``Excited QCD 2012'', Peniche,
Portugal, 7--11 May 2012.}%
}
\author{S. Coito$^*$, G. Rupp
\address{
Centro de F\'isica das Interac\c c\~oes Fundamentais, Instituto Superior
T\'ecnico, Technical University of Lisbon, P-1049-001 Lisboa, Portugal}
\and
E. van Beveren
\address{
Centro de F\'{\i}sica Computacional, Departamento de F\'{\i}sica,
Universidade de Coimbra, P-3004-516 Coimbra, Portugal}
}

\maketitle

\begin{abstract}
Because of the controversial $X(3872)$ meson's very close proximity to the
$D^0\overline{D}^{*0}$ threshold, this charmonium-like resonance is often
considered a meson-meson molecule. However, a molecular wave function must be
essentially of a meson-meson type, viz.\ $D^0\overline{D}^{*0}$ in this case,
with no other significant components. We address this issue by employing a
simple two-channel Schr\"{o}dinger model, in which the $J^{PC}=1^{++}$
$c\bar{c}$ and $D^0\overline{D}^{*0}$ channels can communicate via the
$^{3\!}P_0$ mechanism, mimicked by string breaking at a sharp distance $a$.
Thus, wave functions and their probabilities are computed, for different
bound-state pole positions approaching the $D^0\overline{D}^{*0}$ threshold
from below. We conclude that at the PDG $X(3872)$ mass and for reasonable
values of $a$, viz.\ $2.0$--$3.0$ GeV$^{-1}$, the $c\bar{c}$ component remains
quite substantial and certainly not negligible, despite accounting for only
about 6--10\% of the total wave-function probability, owing to the
naturally long tail of the $D^0\overline{D}^{*0}$ component.
\end{abstract}

\mbox{ } \\ 		
The $X(3872)$ charmonium-like meson is by now a very well established
resonance \cite{PRD86p010001}. It was first observed in 2003, by the Belle
Collaboration \cite{PRL91p262001}, in the decay
$B^{\pm} \rightarrow K^{\pm} \pi^+\pi^- J\!/\!\psi$, with significance in excess
of 10$\sigma$. Since then, it has been confirmed by several collaborations,
viz.\ Belle, BaBar, CDF, D0, CLEO, and, more recently, by LHCb (see the 2012
PDG \cite{PRD86p010001} listings for references). The PDG summary table lists
the $X(3872)$ as an isoscalar state with positive $C$-parity, from the observed
$\gamma J\!/\!\psi$ decay, but unknown $J$ and $P$, having an average mass
$m=3871.68\pm0.17$ MeV/$c^2$ and a width $\Gamma<1.2$ MeV/$c^2$. The two most
likely $J^{PC}$ assignments are $1^{++}$ and $2^{-+}$, while the observed 
hadronic decay modes are $\rho^0 J\!/\!\psi$, $\omega J\!/\!\psi$,
$D^0\overline{D}^{*0}$, and $D^0\overline{D}^{0}\pi^0$. Henceforth, we shall
denote $D^0\overline{D}^{*0}$ simply by $D^0D^{*0}$.

Meson spectroscopists have been puzzled  by the $X(3872)$, 
because of its low mass as compared to predictions of conventional
quark models, as well as its remarkable proximity to the $D^0D^{*0}$
threshold, being ``bound'' by only 0.15 MeV \cite{PRD86p010001}.
This has led to a plethora of model descriptions of the $X(3872)$, viz.\
as a $c\bar{c}$ state, meson-meson (MM) molecule, tetraquark, or hybrid meson.
For a number of reviews on the many different approaches and the 
experimental situation, see \cite{PR429p243}. Recently, we have described
\cite{EPJC71p1762} the $X(3872)$ as a regular but ``unquenched'' $1^{++}$
(\tpo) charmonium meson, whose physical mass is dynamically shifted about
100 MeV downwards from the bare \ttpo\ $c\bar{c}$ state due to its strong
coupling to the $S$-wave $D^0D^{*0}$ and $D^\pm D^{*\mp}$ channels, besides
several other OZI-allowed and OZI-forbidden
($\rho^0 J\!/\!\psi$, $\omega J\!/\!\psi$) channels. Thus, the observed
hadronic $X(3872)$ properties were well reproduced \cite{EPJC71p1762}.

Nevertheless, the closeness of the $X(3872)$ to the $D^0D^{*0}$ threshold
seems to favour a molecular interpretation \cite{PRD76p094028}. In the latter
paper, it is stated that, whatever the original mechanism generating the
resonance, a near-threshold bound state will always have a molecular structure.
This implies that the MM component of the wave function, i.e., 
$D^0D^{*0}$, should be the only relevant one. Here, we shall study this issue
in a simplified, coordinate-space version of the model employed in
\cite{EPJC71p1762}, restricted to the most important channels, viz.\
$c\bar{c}$ and $D^0D^{*0}$. Note that, even if the $X(3872)$ is essentially a
molecule, it will mix with $c\bar{c}$ states having the same quantum
numbers.

Now we turn to the two-channel model used in \cite{ZPC19p275}, with parameters
adjusted for the $X(3872)$. Consider a coupled $q\bar{q}\,$-$\,M_1M_2$
system, with the $q\bar{q}$ pair confined through a harmonic-oscilator (HO)
potential, whereas the two mesons $M_1,M_2$ are free. The correponding
$2\times2$ radial Schr\"odinger equation is given by Eq.~(\ref{schr}), with the
Hamiltonians (\ref{hc}) and (\ref{hf}). Here, $\mu_{c,f}$ is
the reduced mass in either channel, $m_{q}=m_{\bar{q}}$ the constituent
quark mass, $l_c,l_f$ the orbital angular momenta, and $\omega$ the HO
frequency:
\be
\label{schr}
\left(\barr{cc}
h_c & V\\
V & h_f
\earr\right)
\left(\barr{c}
u_c\\
u_f
\earr\right)=
E\left(\barr{c}
u_c\\
u_f
\earr\right) \; ;
\ee
\be
\label{hc}
h_c=\frac{1}{2\mu_c}\bigg(-\frac{d^ 2}{dr^ 2}+\frac{l_c(l_c+1)}{r^2}\bigg)+
\frac{1}{2}\mu_c\omega^2r^2+m_q+m_{\bar{q}} \; ;
\ee
\be
\label{hf}
h_f=\frac{1}{2\mu_f}\bigg(-\frac{d^ 2}{dr^ 2}+\frac{l_f(l_f+1)}{r^2}\bigg)+
M_1+M_2 \; .
\ee
Note that we use here relativistic definitions for the MM reduced
mass $\mu_f$ and relative momentum $k$, even below threshold contrary to
\cite{ZPC19p275}, though this is practically immaterial for the $X(3872)$. 
At some ``string-breaking'' distance $a$, transitions between
the two channels are described by an off-diagonal point-like potential with
strength $g$
\be
\label{pot}
V=\frac{g}{2\mu_ca}\delta(r-a) \; .
\ee
Continuity and twice integrating Eqs.~(\ref{schr}--\ref{hf}) yields the
boundary conditions
\be
\label{bc1}
u_c'(r\upr a)-u_c'(r\dar a)+\frac{\lambda}{a}u_f(a) \, = \, 
u_f'(r\upr a)-u_f'(r\dar a)+\frac{\lambda\mu_f}{a\mu_c}u_c(a)=0 \; ,
\ee
\be
\label{bc2}
u_c(r\upr a)=u_c(r\dar a) \;\;\; \mbox{and} \;\;\;
u_f(r\upr a)=u_f(r\dar a) \; .
\ee\mbox{} \\[2mm]
A general solution to this problem is given by Eqs.~(\ref{wfc}) and
(\ref{fwf}) for the confined and the MM state, respectively. The
two-component function $u(r)=(u_c(r),u_f(r))$ is related to the radial
wave function as $u(r)=rR(r)$:
\be
\label{wfc}
u_c(r)=
\left\lbrace\barr{lc}
A_cF_c(r) & r<a \; , \\[1mm]
B_cG_c(r) & r>a \; ;
\earr\right.
\ee
\be
\label{fwf}
u_f(r)=
\left\lbrace\barr{lc}
A_f J_{l_f}(kr)& r<a \; , \\[2mm]
B_f\Big\lbrack J_{l_f}(kr)k^{2l_f+1}\cot\big(
\delta_{l_f}(E)\big)-N_{l_f}(kr)\Big\rbrack  & r>a \; .
\earr\right.
\ee
Now, $F_c(r)$ vanishes at the origin and $G_c(r)$ falls off
exponentially for $r\to\infty$. Defining then $z=\mu\omega r^2$ and
\be
\label{nu}
\nu=\frac{E-2m_c}{2\omega}-\frac{l_c+3/2}{2} \; ,
\ee
 we get
\be
\label{ffc}
F(r)=\frac{1}{\Gamma(l+3/2)}z^{(l+1)/2}e^{-z/2}\phi(-\nu,l+3/2,z) \; ,
\ee
\be
\label{fgc}
G(r)=-\frac{1}{2}\Gamma(-\nu)rz^{l/2}e^{-z/2}\psi(-\nu,l+3/2,z) \; .
\ee
Here, the functions $\phi$ and $\psi$ are the confluent hypergeometric
functions of first and second kind, respectively, and the $\Gamma$
function acts as a normalising function. 
The functions $J$ and $N$ in Eq.~(\ref{fwf}) are defined in terms of
the spherical Bessel and Neumann functions $j,n$, i.e.,
$J_l(kr)=k^{-l}rj_l(kr)$ and $N_l(kr)=k^{l+1}rn_l(kr)$.
From the boundary conditions (\ref{bc1},\ref{bc2}) and the explicit
wave-function expressions in Eqs.~(\ref{wfc},\ref{fwf}), we obtain
\be
\label{bc1b}
\barr{l}
G_c'(a)F_c(a)-F_c'(a)G_c(a)=\frac{g}{a}J_{l_f}(ka)F_c(a)\frac{A_f}{B_c} \; , \\
\mbox{}\\
J_{l_f}'(ka)N_{l_f}(ka)-J_{l_f}(ka)N_{l_f}'(ka)=
\frac{g}{a}\frac{\mu_f}{\mu_c}J_{l_f}(ka)F_c(a)\frac{A_c}{B_f} \; .
\earr
\ee
Using next the Wronskian relations 
\bea
\label{bc2b}
W(F_c(a),G_c(a))&=&
\dsp\lim_{r\to a}\left[F_c(r)G_c'(r)-F_c'(r)G_c(r)\right]=1\;,\\
W(N_{l_f}(ka),J_{l_f}(ka))&=&\dsp\lim_{r\to a}\left[N_{l_f}(kr)J_{l_f}'(kr) -
N_{l_f}'(kr)J_{l_f}(kr)\right]=-1 \; . \nonumber
\eea
yields
\be
\label{ampr1}
A_fB_f=-\frac{\mu_f}{\mu_c}A_cB_c
\ee
and
\be
\label{ampr2}
\frac{A_f}{B_f}=-\bigg[\frac{g^2}{a^2}
\frac{\mu_f}{\mu_c}J_{l_f}^2(ka)F_c^2(a)\bigg]^{-1}\frac{B_c}{A_c} \; .
\ee
Finally, with the expression for the MM scattering wave function $u_f(r)$
(second line in Eq.~(\ref{fwf})), the final result for
$\cot\delta_{l_f}(E)$ is obtained, reading
\be
\label{cotan}
\cot\big(\delta_{l_f}(E)\big)=-\bigg[g^2\frac{\mu_f}{\mu_c}kj_{l_f}^2(ka)
F_c(a)G_c(a)\bigg]^{-1}+\frac{n_{l_f}(ka)}{j_{l_f}(ka)} \; .
\ee
Now, in the present $X(3872)$ model, there is only one scattering
channel, viz.\ for the $D^0D^{*0}$ system. Thus, poles in the $S$-matrix,
which represent possible resonances, bound states, or virtual states, are
given by the simple relation $\cot\delta_{l_f}(E)=i$. On the other hand,
the solutions to the two-component radial wave function (\ref{wfc},\ref{fwf})
are then fully determined by relations (\ref{ampr1}) and (\ref{ampr2}), up
to an overall normalisation constant.

\begin{figure}[!hb]
\centering
\caption{Dynamical real (solid) and virtual (dashed) pole trajectories
 for $a=2.0$ GeV$^{-1}$ (left) and $a=3.0$ GeV$^{-1}$ (right). The arrows
indicate pole movement for increasing $g$. The PDG \cite{PRD86p010001}
$X(3872)$ mass is labelled by \boldmath$\ast$. Also see Table~\ref{tab:coup}.}
\begin{tabular}{lr}
\hspace*{-50pt}\resizebox{!}{350pt}{\includegraphics{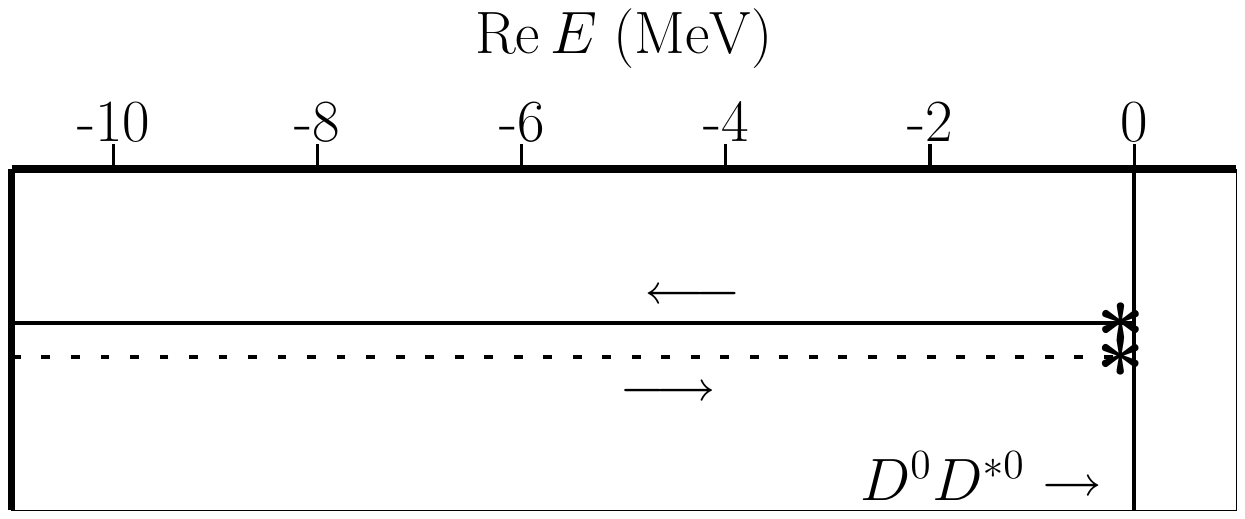}}
&
\hspace*{-90pt}\resizebox{!}{350pt}{\includegraphics{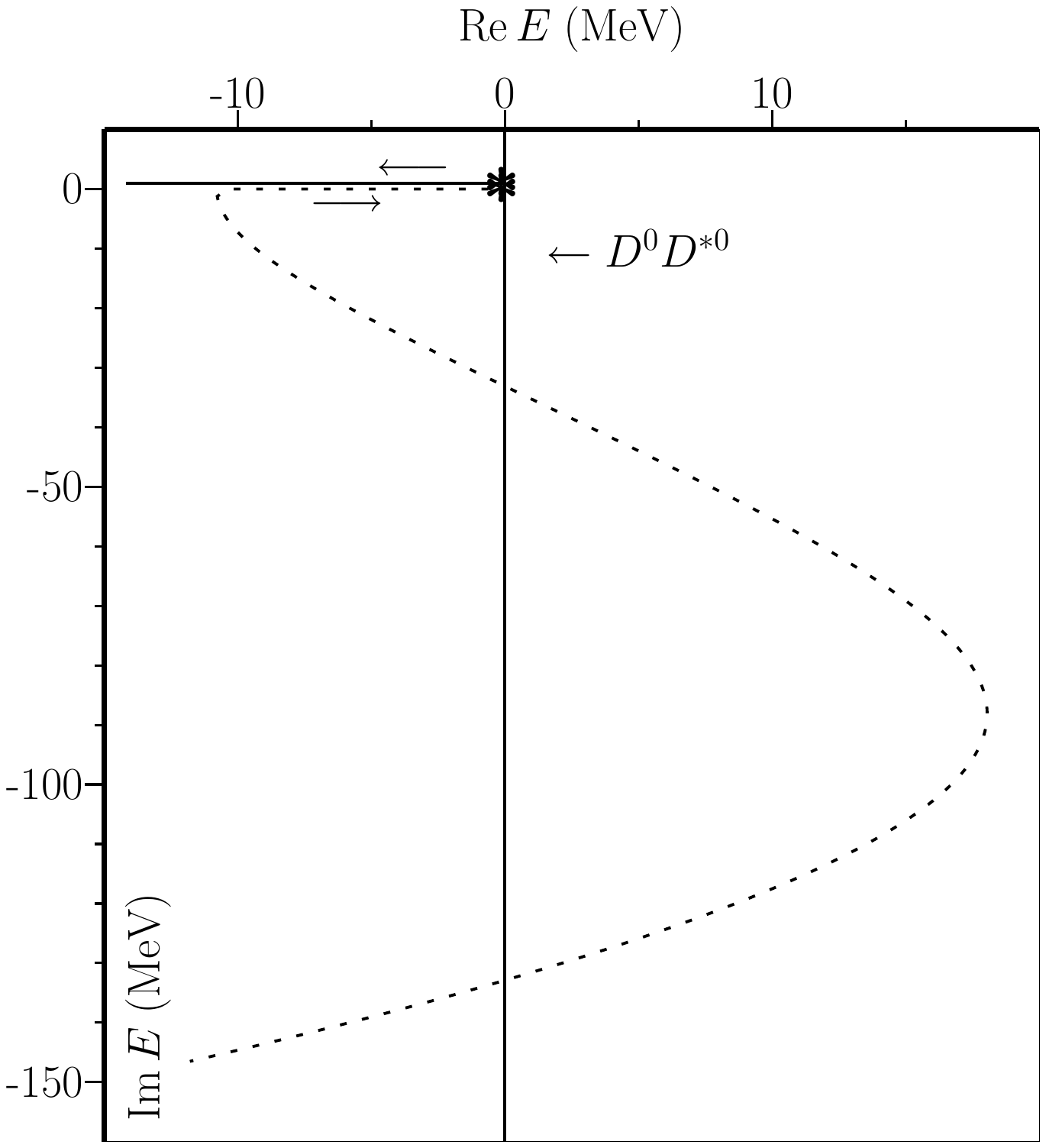}}\\
\end{tabular}
\mbox{ } \\[-5.5cm]
\label{fig:traj}
\end{figure}

Next we apply this formalism to the coupled $c\bar{c}\,$-$\,D^0D^{*0}$ system.
In the confined channel, the $c\bar{c}$ system is in a \ttpo\ state, and so
$l_c=1$, whereas the $D^0D^{*0}$ channel has $l_f=0$.
In Table~\ref{tab:param} we give the fixed parameters of the model, with the HO
\begin{table}[hb]
\centering
\caption{Fixed model parameters
\cite{PRD27p1527}
and $D^0D^{*0}$ threshold.}
\begin{tabular}{c|c|c|c|c|c}
\hline
\hline
Parameter& $\omega$&$m_c$&$m_{D^0}$&$m_{D^{*0}}$&$m_{D^0}+m_{D^{*0}}$\\
\hline \mbox{ } \\[-3.7mm]
Value (MeV)&$190$&$1562$&$1864.84$&$2006.97$&$\bf{3871.81}$ \\[-1mm]
\end{tabular}
\label{tab:param}
\end{table}
frequency $\omega$ and the constituent charm mass as in \cite{PRD27p1527}, 
being unaltered ever since. However, the radial quantum number $\nu$ in 
Eq.~(\ref{nu}) varies as a function of the energy, and therefore will
generally be non-integer, becoming even complex for resonance poles. The
parameter that determines such variations is the coupling $g$. In the
uncoupled case, i.e., for $g=0$, one recovers the bare \tpo\ HO spectrum,
with energies $(3599+2n\omega)$~MeV ($n=0,1,2,\ldots$).
The only other free parameter is the string-breaking distance $a$.
Now we try to find $S$-matrix poles as a function of the coupling $g$ and
for two reasonable values of $a$, viz.\ $2.0$ and $3.0$ GeV$^{-1}$
($\approx\!0.4$ and $0.6$ fm). Searching near the $D^0D^{*0}$ threshold,
a dynamical pole is found, either on the first Riemann sheet, corresponding
to a bound state, or on the second one, which represents a virtual state
(see Ref.~\cite{EPJC71p1762}, second paper). These results are presented
in Table~\ref{tab:coup} and Fig.~\ref{fig:traj}.
\begin{table}[!t]
\centering
\caption{Bound and virtual states near the $D^0D^{*0}$ threshold.}
\mbox{ }\\[-3mm]
\begin{tabular}{c|c|c|c}
\hline
\hline
\mbox{ }\\[-3.5mm]
$\bf{a}$ (GeV$^{-1}$)& $\bf{g}$ & Pole (MeV) & Type of bound state\\
\hline
$2.0$ & $1.133$ & $3871.68$&virtual\\
$2.0$ & $1.150$ & $3871.81$&virtual\\
$2.0$ & $1.153$ & $3871.81$&real\\
$2.0$ & $1.170$ & $3871.68$&real\\
\hline
$3.0$ & $2.097$ & $3871.68$&virtual\\
$3.0$ & $2.144$ & $3871.81$&virtual\\
$3.0$ & $2.150$ & $3871.81$&real\\
$3.0$ & $2.199$ & $3871.68$&real \\[-4mm]
\end{tabular}
\label{tab:coup}
\end{table}

Note that the dynamical pole arises from the $D^0D^{*0}$ continuum and is
not connected to the bare \ttpo\ $c\bar{c}$ state at 3979~MeV, contrary to
the situation in \cite{EPJC71p1762} (first paper). For our study here,
this is of little consequence.

\begin{figure}[t]
\centering
\caption{Radial wave-functions for $E=3871.68$ MeV and $g=1.170$,
$g=2.199$ for $a=2.0$ GeV$^{-1}$ (left) and $a=3.0$ GeV$^{-1}$ (right).
Also see Table~\ref{tab:coup}.}
\mbox{} \\[-10mm]
\begin{tabular}{lr}
\hspace*{-40pt}\resizebox{!}{350pt}{\includegraphics{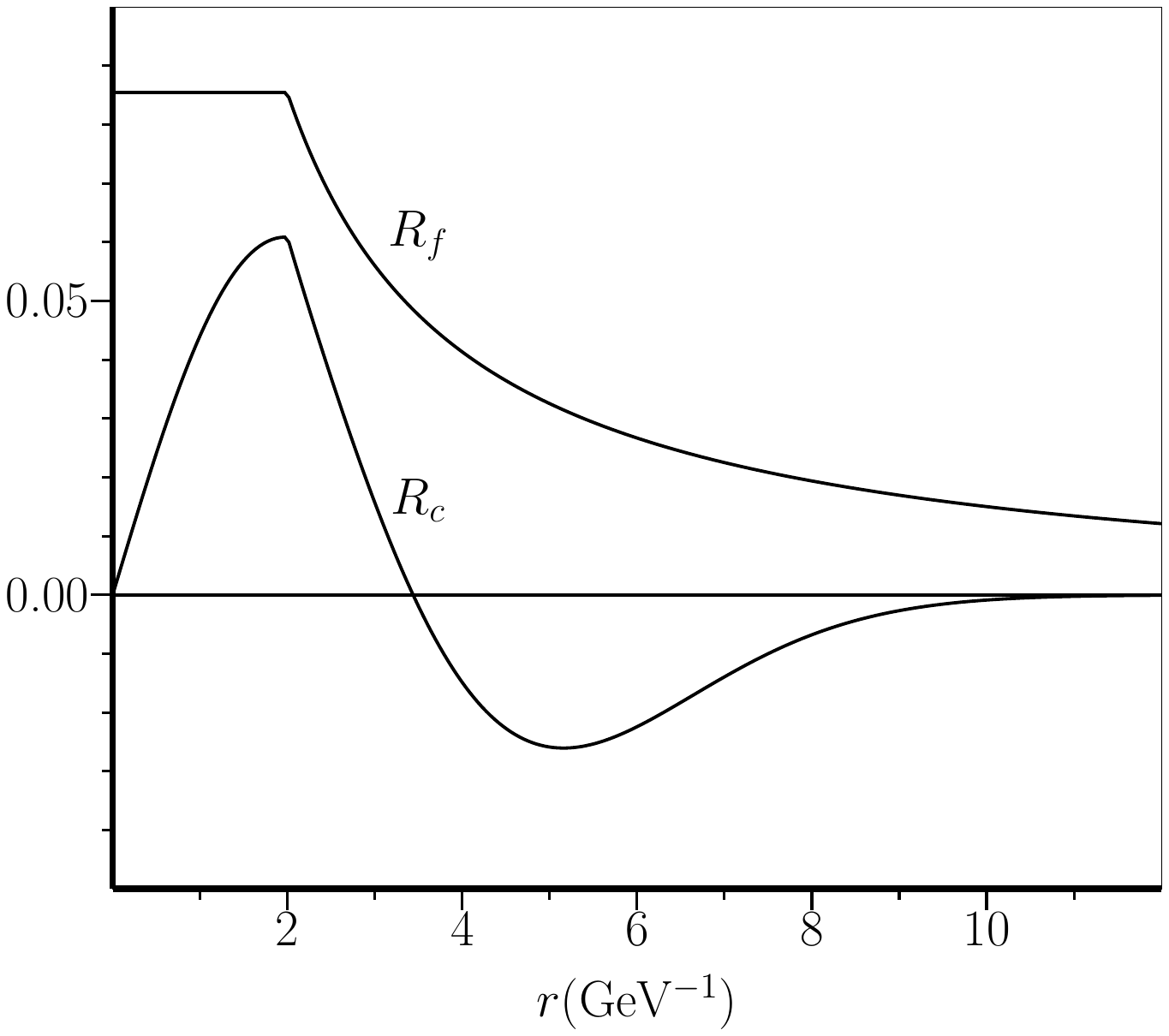}}&
\hspace*{-95pt}\resizebox{!}{350pt}{\includegraphics{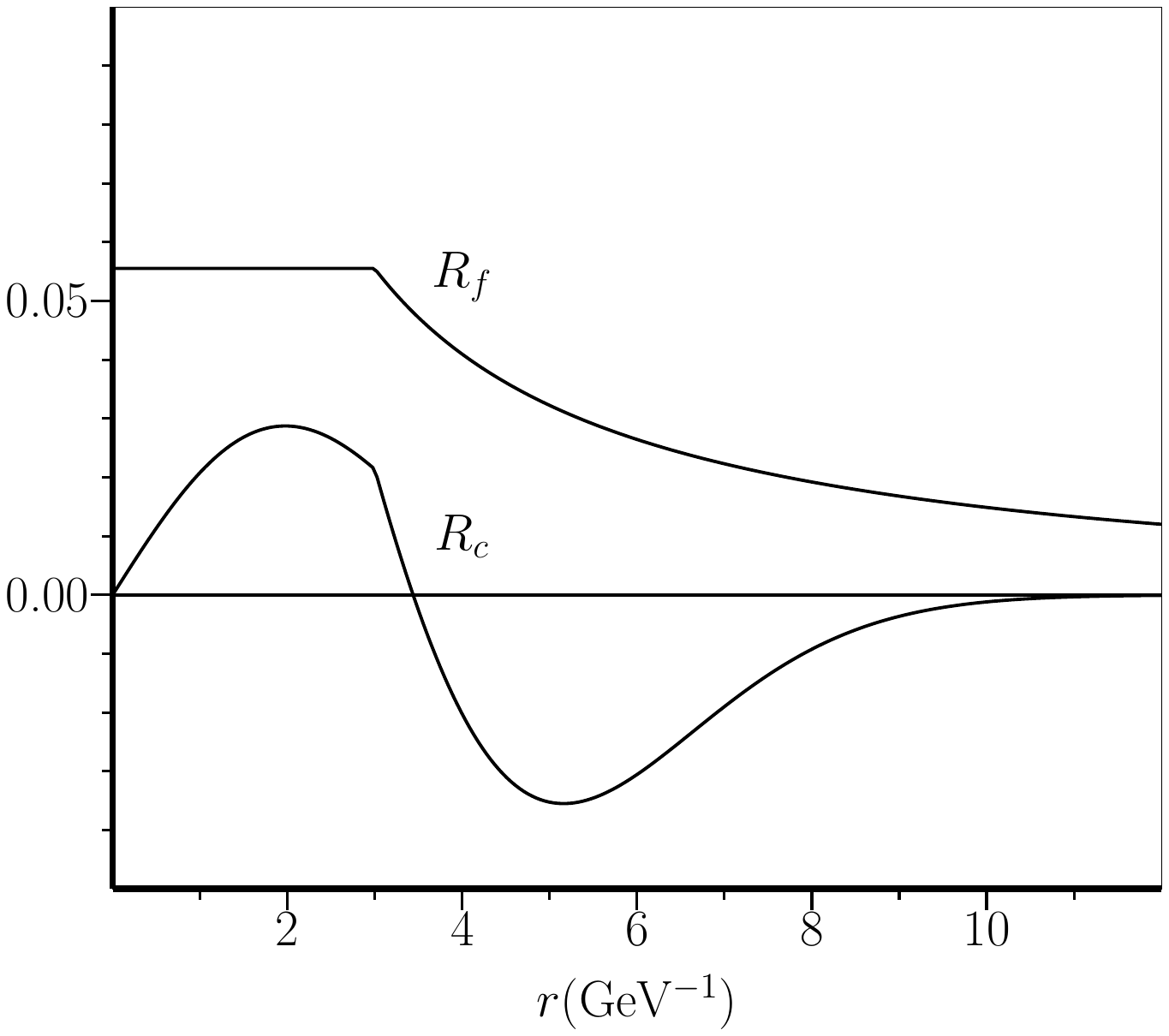}}\\
\mbox{} \\[-70mm]
\end{tabular}
\label{fig:rwf}
\end{figure}

Finally, we depict the normalised two-component wave-function $R(r)$ in
Fig.~\ref{fig:rwf}, evaluated for the PDG \cite{PRD86p010001} $X(3872)$
mass of 3872.68~MeV. One clearly sees the $P$-wave behaviour of the
$c\bar{c}$ component, whereas the $D^0D^{*0}$ is in an $S$-wave.
Moreover, the $c\bar{c}$ admixture is certainly not negligible, despite
the low total probablities of 6.13\% and 10.20\%, for $a\!=\!2$ GeV$^{-1}$
and $a\!=\!3$ GeV$^{-1}$, respectively, which are logical because of the
very long tail of the $D^0D^{*0}$ component; also see \cite{12064877}. Soon
we will publish more detailed work.


\begin{thebibliography}{12}

\bibitem{PRD86p010001}
J.~Beringer {\it et al.} [Particle Data Group], Phys.\ Rev.\ D {\bf 86}
(2012) 010001. 

\bibitem{PRL91p262001}
S.-K.~Choi {\it et al.}  [The Belle Collaboration],
Phys.\ Rev.\ Lett.\  {\bf 91} (2003) 262001.

\bibitem{PR429p243}
E.~S.~Swanson,
Phys.\ Rept.\ {\bf429}, 243 (2006); 
E.~Klempt amd A.~Zaitsev,
Phys.\ Rept.\ {\bf454} (2007) 1; 
K.~Seth, 
Prog.\ in Part.\ and Nucl.\ Phys.\ {\bf67} (2012) 390;
J.~Zhang,
arXiv:1112.0841 [hep-ex].

\bibitem{EPJC71p1762}
S.\ Coito, G.\ Rupp, and E.\ van Beveren, Eur.\ Phys.\ J.\ C {\bf71}
(2011) 1762;
S.\ Coito, G.\ Rupp, and E.\ van Beveren, Acta Phys.\ Polon.\ Suppl.\ {\bf3}
(2010) 983.

\bibitem{PRD76p094028}
Eric Braaten and Meng Lu, Phys.\ Rev.\ D {\bf76} (2007) 094028.

\bibitem{ZPC19p275}
E.~van Beveren, C. Dullemond, and T.A. Rijken, Z. Phys. C {\bf19} (1983) 275.

\bibitem{PRD27p1527}
E.~van Beveren, G.~Rupp, T.~A.~Rijken, and C.~Dullemond,
Phys.\ Rev.\  D {\bf 27} (1983) 1527.

\bibitem{12064877}
M.~Takizawa and S.~Takeuchi,
arXiv:1206.4877 [hep-ph].

\end{thebibliography}
\end{document}